\documentclass[11pt]{article}
\usepackage{graphicx}
\usepackage{tabularx}
\usepackage{url}
\usepackage{amsthm}
\usepackage{amsmath}
\usepackage{amsfonts}

\theoremstyle{plain}
\newtheorem{theorem}{Theorem}
\newtheorem{lemma}{Lemma}

\theoremstyle{definition}

\theoremstyle{remark}
\newtheorem{remark}{Remark}

\date{}

\begin{document}

\title{Diagnosis of Constant Faults in Switching Networks}
\author{Mikhail Moshkov\thanks{Computer, Electrical and Mathematical Sciences \& Engineering Division and Computational Bioscience Research Center,
King Abdullah University of Science and Technology (KAUST),
Thuwal 23955-6900, Saudi Arabia. Email: mikhail.moshkov@kaust.edu.sa.
}}
\maketitle

\begin{abstract}
In this paper, we study decision trees for diagnosis of constant faults in
switching networks. Each constant fault consists in assigning Boolean constants to
some edges of the network instead of literals. The problem of diagnosis is to
recognize the function implemented by the switching network with a constant fault
from a given set of faults. For this problem solving, we use decision trees. Each
query (attribute) of a decision tree consists of observing the value of function
implemented by the faulty switching network on a given tuple of variable values.
We study the depth of decision trees for diagnosis of arbitrary and specially
constructed switching networks and the complexity of diagnostic decision tree
construction.
\end{abstract}


\section{Introduction}

Switching networks are among most fundamental models for computing Boolean
functions \cite{Jukna12}. In this paper, we study decision trees for diagnosis
of constant faults in these networks.

A switching network is an undirected graph with two
poles in which literals (variables or negations of variables) are assigned
to edges. For a given tuple of variable values, the value of the function
implemented by the network is equal to $1$ if and only if there exists a
path between poles such that the value of each literal assigned to the edges
in this path is equal to $1$.
We consider faults each of which consists in
assigning Boolean constants to some edges of the network instead of literals. The problem
of diagnosis is to recognize the function implemented by the switching
network with a constant fault from a given set of faults.

There are two types of algorithms for this problem solving: non-adaptive
(diagnostic tests) and adaptive (diagnostic decision trees). The most part of papers
related to the diagnosis of switching networks is devoted to the study of diagnostic
tests -- see, for example, \cite{Chegis58,Madatyan70,Red'kin19}. There is only a small number
of papers considering decision trees for diagnosis of switching networks
\cite{BusbaitCHM15,BusbaitM16,Goduhina88}. In these papers, iteration-free switching
networks are studied in which edges are labeled with pairwise different variables.

In this paper, we study decision trees for diagnosis of arbitrary switching networks.
Each query (attribute) of a decision
tree consists of observing the value of function implemented by the faulty
switching network  on a given tuple of variable values.

Let $P$ be a switching network. We denote by $L(P)$ the number of edges in $P$.
We consider three types of constant faults: $%
\{0,1\}$-faults, $\{0\}$-faults, and $\{1\}$-faults of $P$. Let $C\in
\{\{0,1\},\{0\},\{1\}\}$. A $C$-fault of $P$ consists in assigning of
constants from $C$ to some edges of $P$ instead of literals. Let $R$ be a
set of $C$-faults of $P$. The problem of diagnosis of $C$-faults for $P$
and $R$: for the switching network $P$ with a $C$-fault $\rho $ from $R$, we
should find a $C$-fault $\delta $ of $P$ such that the network $P$ with the
fault $\delta $ implements the same function as the network $P$ with the
fault $\rho $. We denote by $h_{C}(P)$ the minimum depth of a decision tree,
which solves the problem of diagnosis of $C$-faults for $P$ and the set of
all possible $C$-faults of $P$.

We obtain a number of negative results related to the
 decision trees for diagnosis of arbitrary switching networks.

First, we show
that there are simple switching networks for which the depth of decision
trees for diagnosis of constant faults is high. We prove that, for any $C\in
\{\{0,1\},\{0\},\{1\}\}$ and any natural $n$, there exists a switching
network $P$  such that $L(P)=2n$ and $h_{C}(P)\geq
2^{n}$.

Then we show that there are simple switching networks such that, for all switching networks implementing the same functions, the depth of decision
trees for diagnosis of constant faults is high.
We  prove that, for any $C\in \{\{0,1\},\{0\},\{1\}\}$ and any natural $n$, there
exists a switching network $P$ such that $%
L(P)=n^{2}+n$ and $h_{C}(Q)\geq \binom{n}{\left\lceil \frac{n}{2}%
\right\rceil }$ for any switching network $Q$ implementing the same function
as $P$.

For any $C\in \{\{0,1\},\{0\},\{1\}\}$, we define the algorithmic problem $%
Con(C)$: for a given switching network $P$ with $L(P)\geq 1$ and a set $R$
of $C$-faults of $P$, it is required to construct a decision tree $\Gamma $,
which solves the problem of diagnosis of $C$-faults for $P$ and $R$. One can
show that there exists a decision tree, which solves this problem and has at
most $2\left\vert R\right\vert -1$ nodes. We prove that, for any $C\in
\{\{0,1\},\{0\},\{1\}\}$, the problem $Con(C)$ is NP-hard.

The remainder of this paper is organized as follows. In Section \ref{S8.1.1},
main notions are considered. In Section \ref{S8.1.2}, switching networks for symmetric
Boolean functions are discussed. In Section \ref{S8.1.3}, we study the depth of
decision trees for diagnosis of arbitrary and specially constructed switching
networks. In Section \ref{S8.1.4}, the complexity of diagnostic decision tree
construction is considered. Section \ref{S8.1.5} contains short conclusions.

\section{Main Notions \label{S8.1.1}}

In this section, we consider main notions related to switching networks,
constant faults, and decision trees for fault diagnosis. We also prove a
simple lower bound on the depth of decision trees for fault diagnosis.

A \emph{switching network }is an undirected connected graph $P$ with multiple edges
and without loops in which two different nodes called \emph{poles} are
fixed. Each edge of $P$ is labeled with a literal from the set $\{x_{i},\bar{%
x}_{i}:i\in \{0,1,2,\ldots \}\}$, where $x_{i}$ is a variable, and $\bar{x}%
_{i}$ is its negation. We denote by $L(P)$ the number of edges in the
switching network $P$.

We consider three types of constant faults: $\{0,1\}$-faults, $\{0\}$%
-faults, and $\{1\}$-faults of $P$. Let $C\in \{\{0,1\},\{0\},\{1\}\}$. A $C$%
-fault of $P$ consists in assigning of constants from $C$ to some edges of $%
P $ instead of literals.

Let $\{x_{i_{1}},\ldots ,x_{i_{m}}\}$ be the set of all variables from
literals attached to edges of $P$. We will say that $x_{i_{1}},\ldots
,x_{i_{m}}$ are \emph{input variables }of $P$. We correspond to the
switching network $P$ and its $C$-fault $\rho $ a Boolean function $%
f_{P,\rho }(x_{i_{1}},\ldots ,x_{i_{m}})$ that is implemented by the
switching network $P$ with the fault $\rho $. Let $\xi $ be a \emph{simple}
path (without repeating nodes) between poles of $P$. We denote by $c_{\rho
}(\xi )$ the conjunction of all Boolean functions (literals or constants)
attached to edges of the path $\xi $ in the switching network $P$ with the
fault $\rho $. Then $f_{P,\rho }(x_{i_{1}},\ldots ,x_{i_{m}})$ is the
disjunction of all conjunctions $c_{\rho }(\xi )$ corresponding to simple
paths $\xi $ between poles of $P$. We denote $f_{P}=$ $f_{P,\lambda }$,
where $\lambda $ is the \emph{empty} fault (no constants are assigned to
edges of $P$), and will say that the network $P$ \emph{implements }%
the function $f_{P}$.

Let $R$ be a set of $C$-faults of $P$. The \emph{problem of diagnosis of }$%
C $-\emph{faults for} $P$ \emph{and} $R$: for the switching network $P$ with a $C$%
-fault $\rho $ from $R$, we should find a $C$-fault $\delta $ of $P$ such
that $f_{P,\rho }=f_{P,\delta }$. To resolve this problem, we can ask about
values of the function $f_{P,\rho }$ on arbitrary tuples from $\{0,1\}^{m}$.

As algorithms for solving the problem of diagnosis of $C$-faults, we will
consider \emph{decision trees} each of which is a directed tree with root.
Terminal nodes of the tree are labeled with $C$-faults of $P$. Each
nonterminal node is labeled with an $m$-tuple from $\{0,1\}^{m}$. Two edges
start in this node that are labeled with $0$ and $1$, respectively. The
\emph{depth }$h(\Gamma )$ of a decision tree $\Gamma $ is the maximum length
of a path from the root to a terminal node.

Let $\Gamma $ be a decision tree. For the switching network $P$ with a $C$%
-fault $\rho $, this tree works in the following way. If the root of $\Gamma
$ is a terminal node, then the output of $\Gamma $ is the fault attached to
the root. Otherwise, we find the value of the function $f_{P,\rho }$ on the $%
m$-tuple attached to the root and pass along the edge, which starts in the
root and is labeled with this value, etc., until we reach a terminal node.
The fault attached to this node is the output of $\Gamma $. We will say that
$\Gamma $ \emph{solves} the problem of diagnosis of $C$-faults for $P$ and $%
R $ if, for any $C$-fault $\rho $ from $R$, the output of $\Gamma $ is a $C$%
-fault $\delta $ of $P$ such that $f_{P,\rho }=f_{P,\delta }$. We denote by $%
h_{C}(P)$ the minimum depth of a decision tree, which solves the problem of
diagnosis of $C$-faults for $P$ and the set of all $C$-faults of $P$
including the empty fault $\lambda $, i.e., the problem of diagnosis of
arbitrary $C$-faults of $P$.

We consider now a lower bound on $h_{C}(P)$.

\begin{lemma}
\label{L8.1.1}Let $C\in \{\{0,1\},\{0\},\{1\}\}$, $P$ be a switching network
with $m$ input variables, $\rho _{0},\rho _{1},\ldots ,\rho _{r}$ be $C$%
-faults of $P$ such that $f_{P,\rho _{0}}\neq f_{P,\rho _{i}}$ for $%
i=1,\ldots ,r$, and $t$ be the minimum cardinality of a set of tuples from $%
\{0,1\}^{m}$ on which the function $f_{P,\rho _{0}}$ is different from the
functions $f_{P,\rho _{1}},\ldots ,f_{P,\rho _{r}}$. Then
\begin{equation*}
h_{C}(P)\geq t\;.
\end{equation*}
\end{lemma}

\begin{proof}
Let $\Gamma $ be a decision tree, which solves the problem of diagnosis of
arbitrary $C$-faults for $P$ and for which $h(\Gamma )=h_{C}(P)$. Let the
work of $\Gamma $ on the network $P$ with the fault $\rho _{0}$ finish in
the terminal node $v$ of $\Gamma $. We denote by $\xi $ the path from the
root of $\Gamma $ to the terminal node $v$. It is clear that, on the set of $%
m $-tuples attached to nonterminal nodes of $\xi $, the function $f_{P,\rho
_{0}}$ is different from the functions $f_{P,\rho _{1}},\ldots ,f_{P,\rho
_{r}}$. Therefore the length of $\xi $ is at least $t$. Thus, $h(\Gamma )\geq
t$ and $h_{C}(P)\geq t$.
\end{proof}

\section{Switching Networks for Symmetric Boolean Functions \label{S8.1.2}%
}

\begin{figure}[h]
	\begin{center}
		\includegraphics[width=38mm]{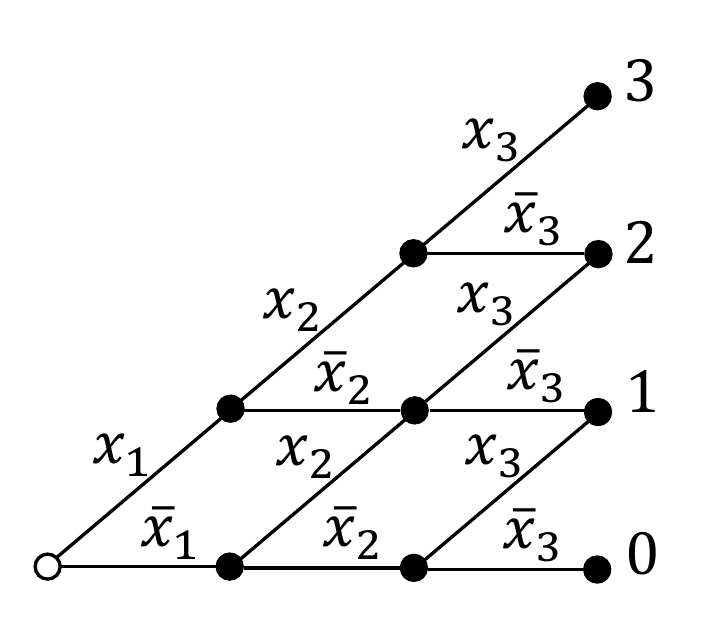}
		\caption{Network $Q$}
		\label{fig8.1}
	\end{center}
\end{figure}

A Boolean function $f(x_{1},\ldots ,x_{n})$ is called \emph{symmetric} if
its value depends only on the number of ones in the input. This function can
be represented by the tuple $\tilde{t}=(t_{0},t_{1},\ldots ,t_{n})\in
\{0,1\}^{n+1}$, where, for $i=0,1,\ldots ,n$, $t_{i}$ is the value of the
function $f$ on each tuple with exactly $i$ ones. Shannon in \cite{Shannon38}
proposed simple switching network $P_{\tilde{t}}$ with $L(P_{\tilde{t}%
})=n^{2}+n$ implementing the symmetric function $f$ that does not equal to $%
0 $ identically. The network $Q$ depicted in Fig. \ref{fig8.1} illustrates the case, when $%
n=3$ (here and below in the figures the poles are unpainted nodes). Let $t_{i_{1}},\ldots ,t_{i_{p}}$ be all digits equal to $1$ in the
tuple $\tilde{t}=(t_{0},t_{1},\ldots ,t_{3})$ representing a symmetric
Boolean function $f(x_{1},x_{2},x_{3})$. Transform the network $Q$ into a
switching network $P_{\tilde{t}}$ implementing the function $f$. One pole of
$P_{\tilde{t}}$ is the left-most node of $Q$. The second pole is obtained by
merging of nodes of $Q$ labeled with the numbers $t_{i_{1}},\ldots
,t_{i_{p}} $.

Let $n$ be a natural number. For any natural number $m$ such that $m+1\leq n$, we denote by $\psi _{n}^{\geq m+1}$ the symmetric
Boolean function with $n$ variables $x_{1},\ldots ,x_{n}$ that is equal to $%
1 $ if and only if the number of ones in the input is at least $m+1$. We
denote by $\psi _{n}^{=\left\lceil \frac{n}{2}\right\rceil }$ the symmetric
Boolean function with $n$ variables $x_{1},\ldots ,x_{n}$ that is equal to $%
1 $ if and only if the number of ones in the input is equal to $\left\lceil
\frac{n}{2}\right\rceil $. We denote by $\psi _{n}^{\neq \left\lceil \frac{n%
}{2}\right\rceil }$ the symmetric Boolean function with $n$ variables $%
x_{1},\ldots ,x_{n}$ that is equal to $1$ if and only if the number of ones
in the input does not equal to $\left\lceil \frac{n}{2}\right\rceil $. For
each of these functions, there is a switching network that implements it and
contains $n^{2}+n$ edges. It is clear that this network can be constructed
in a time polynomial on $n$.

\section{Diagnosis of Arbitrary and Specially Constructed Switching
Networks \label{S8.1.3}}

In this section, we consider two negative results related to the depth of
decision trees for diagnosis of switching networks. First, we prove that
there are simple switching networks for which the depth of decision trees
for diagnosis of constant faults is high.

\begin{theorem}
\label{T8.1.1} For any $C\in \{\{0,1\},\{0\},\{1\}\}$ and any natural $n$,
there exists a switching network $P_C$ with $n$ input variables such that $%
L(P_C)=2n$ and $h_{C}(P_C)\geq 2^{n}$.
\end{theorem}

\begin{proof}
(a) Let $C\in \{\{0,1\},\{0\}\}$. As switching networks $P_{\{0,1\}}$ and $P_{\{0\}}$,  we consider the switching network $S_1$
depicted in Fig. \ref{fig8.2}. It is clear that $S_1$ has $n$ input variables and $L(S_1)=2n$%
. We denote by $\rho $ a $\{0\}$-fault of $S_1$ such that the constant $0$ is
assigned to each edge of $S_1$. It is clear that $f_{S_1,\rho }=0$. Let $\tilde{%
\delta}=(\delta _{1},\ldots ,\delta _{n})\in \{0,1\}^{n}$. We now describe a
$\{0\}$-fault $\rho (\tilde{\delta})$ of $S_1$: for $i=1,\ldots ,n$, if $%
\delta _{i}=0$, then we assign the constant $0$ to the edge labeled with $%
x_{i}$, and if $\delta _{i}=1$, then we assign $0$ on the edge labeled with $%
\bar{x}_{i}$. It is clear that $f_{S_1,\rho (\tilde{\delta})}=x_{1}^{\delta
_{1}}\cdots x_{n}^{\delta _{n}}$, where $x^{\delta}=x$ if $\delta=1$ and $x^{\delta}=\bar{x}$ if $\delta=0$. One can show that $f_{S_1,\rho }\neq
f_{S_1,\rho (\tilde{\delta})}$ for any $\tilde{\delta}\in \{0,1\}^{n}$ and $\
2^{n}$ is the minimum cardinality of a set of tuples from $\{0,1\}^{n}$ on
which the function $f_{S_1,\rho }$ is different from the functions $f_{S_1,\rho (%
\tilde{\delta})}$, $\tilde{\delta}\in \{0,1\}^{n}$. It is clear that any $%
\{0\}$-fault of $S_1$ is a $\{0,1\}$-fault of $S_1$. By Lemma \ref{L8.1.1}, $%
h_{C}(S_1)\geq 2^{n}$.

\begin{figure}[h]
	\begin{center}
		\includegraphics[width=55mm]{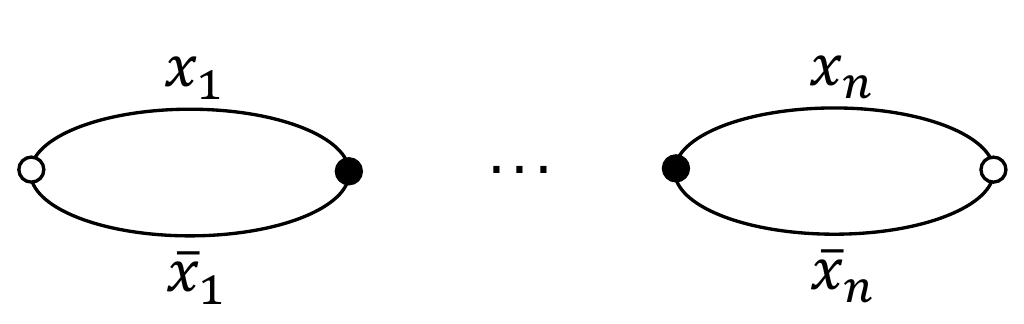}
		\caption{Switching network $S_1$}
		\label{fig8.2}
	\end{center}
\end{figure}

(b) Let $C=\{1\}$. As switching network $P_{\{1\}}$, we consider  the switching network  $S_2$ depicted in Fig. \ref{fig8.3}.
It is clear that $S_2$ has $n$ input variables and $L(S_2)=2n$. It is also clear that
$f_{S_2,\lambda }=0$, where $\lambda $ is the empty fault. Let $\tilde{\delta}%
=(\delta _{1},\ldots ,\delta _{n})\in \{0,1\}^{n}$. We now describe a $\{1\}$%
-fault $\rho (\tilde{\delta})$ of $S_2$: for $i=1,\ldots ,n$, if $\delta
_{i}=0 $, then we assign $1$ to the edge labeled with $x_{i}$, and if $%
\delta _{i}=1 $, then we assign $1$ to the edge labeled with $\bar{x}_{i}$.
It is clear that $f_{S_2,\rho (\tilde{\delta})}=x_{1}^{\delta _{1}}\cdots
x_{n}^{\delta _{n}}$. One can show that $f_{S_2,\lambda }\neq f_{S_2,\rho (\tilde{%
\delta})}$ for any $\tilde{\delta}\in \{0,1\}^{n}$ and $\ 2^{n}$ is the
minimum cardinality of a set of tuples from $\{0,1\}^{n}$ on which the
function $f_{S_2,\lambda }$ is different from the functions $f_{S_2,\rho (\tilde{%
\delta})}$, $\tilde{\delta}\in \{0,1\}^{n}$. By Lemma \ref{L8.1.1}, $%
h_{C}(S_2)\geq 2^{n}$.
\end{proof}

\begin{figure}[h]
	\begin{center}
		\includegraphics[width=62mm]{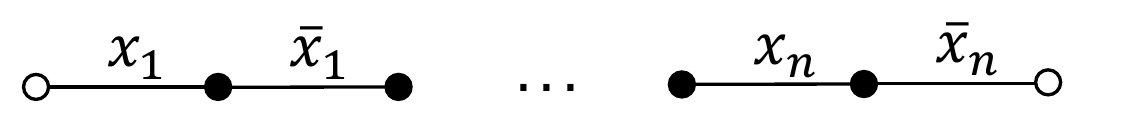}
		\caption{Switching network $S_2$}
		\label{fig8.3}
	\end{center}
\end{figure}

We now show that there are simple switching networks such that, for all switching networks implementing the same functions, the depth of decision
trees for diagnosis of constant faults is high.

\begin{theorem}
\label{T8.1.2} For any $C\in \{\{0,1\},\{0\},\{1\}\}$ and any natural $n$,
there exists a switching network $P$ with $n$ input variables such that $%
L(P)=n^{2}+n$ and $h_{C}(Q)\geq \binom{n}{\left\lceil \frac{n}{2}%
\right\rceil }$ for any switching network $Q$ implementing the same function as $P$.
\end{theorem}

\begin{proof}
(a) Let $C\in \{\{0,1\},\{1\}\}$. We denote by $P$ the switching network
with $n$ input variables $x_{1},\ldots ,x_{n}$, which implements the
symmetric function $\psi =\psi _{n}^{\neq \left\lceil \frac{n}{2}%
\right\rceil }$ and for which $L(P)=n^{2}+n$ (see Sect. \ref{S8.1.2}).

Let $Q$ be a switching network with $n$ input variables $x_{1},\ldots ,x_{n}$%
, which implements the function $\psi $. Let $\tilde{\delta}=(\delta
_{1},\ldots ,\delta _{n})$ be a tuple from the set $\{0,1\}^{n}$ containing
exactly $k=\left\lceil \frac{n}{2}\right\rceil $ ones. Then $\psi (\tilde{\delta})=0$. For definiteness, we
assume that $\tilde{\delta}=(1,\ldots ,1,0,\ldots ,0)$. We know that $\psi $
is equal to the disjunction of all conjunctions $c_{\lambda }(\xi )$
corresponding to simple paths $\xi $ between poles of $Q$, where $\lambda $
is the empty fault of $Q$ (see Sect. \ref{S8.1.1}). For the tuple $\tilde{\delta}$, each conjunction
is equal to $0$. Therefore, in each conjunction, there is at least one
literal from the set $A=\{\bar{x}_{1},\ldots ,\bar{x}_{k},x_{k+1},\ldots
,x_{n}\}$.

Denote $B=\{x_{1},\ldots ,x_{k},\bar{x}_{k+1},\ldots ,\bar{x}%
_{n}\}$.
Let $\rho (\tilde{\delta})$ be the fault of $Q$ obtained by
assigning the constant $1$ to each edge labeled with a literal from $B$. The
switching network $Q$ with the fault $\rho (\tilde{\delta})$ implements the
function $f_{Q,\rho (\tilde{\delta})}$. This function is the disjunction of
all conjunctions $c_{\rho (\tilde{\delta})}(\xi )$ corresponding to simple
paths $\xi $ between poles of $Q$. Since each conjunction $c_{\lambda }(\xi )$
contains a literal from $A$, each conjunction $c_{\rho (\tilde{\delta}) }(\xi )$
contains a literal from $A$.
Thus, the value of $f_{Q,\rho (\tilde{\delta})}$ on
the tuple $\tilde{\delta}$ is equal to $0$.

Let $i\in \{1,\ldots ,n\}$ and $\tilde{\sigma}_{i}$ be the
tuple obtained from $\tilde{\delta}$ by changing the $i$th digit $\delta
_{i} $ to $\bar{\delta}_{i}$. Since $\psi $ is equal to $1$ on the tuple $%
\tilde{\sigma}_{i}$, the function $f_{Q,\rho (\tilde{\delta})}$ is equal to $%
1$ on the tuple $\tilde{\sigma}_{i}$. Therefore in $Q$ there is a simple
path $\xi $ that contains only literals from $A$ that are based on the variable $%
x_{i} $: $\bar{x}_{i}$ if $i\in \{1,\ldots ,k\}$ and $x_{i}$ if $i\in
\{k+1,\ldots ,n\} $. For this path, $c_{\rho (\tilde{\delta})}(\xi )$ is equal
to $\bar{x}_{i}$ if $i\in \{1,\ldots ,k\}$ and to $x_{i}$ if $i\in \{k+1,\ldots ,n\}$.

As a result, we obtain $%
f_{Q,\rho (\tilde{\delta})}=$ $\bar{x}_{1}\vee \cdots \vee \bar{x}_{k}\vee
x_{k+1}\vee \cdots \vee x_{n}\vee C_{1}\vee \cdots \vee C_{m}$, where $%
C_{1},\ldots ,C_{m}$ are elementary conjunctions constructed from literals
from $A$ and containing at least two literals each. Let $j\in \{1,\ldots
,m\}$. Then $C_{j}$ can be represented as $aD$, where $a\in A$ and $D$ is an
elementary conjunction with literals from $A$ that does not contain $a$. It is clear that $a\vee aD=a$. As a result, we have $f_{Q,\rho (\tilde{\delta}%
)}=$ $\bar{x}_{1}\vee \cdots \vee \bar{x}_{k}\vee x_{k+1}\vee \cdots \vee
x_{n}$.

Thus, for any tuple $(b_{1},\ldots ,b_{n})\in \{0,1\}^{n}$ with exactly $k$
zeros, $x_{1}^{b_{1}}\vee \cdots \vee x_{n}^{b_{n}}$ is a function
implemented by the switching network $Q$ with some $\{1\}$-fault. It is
clear that the function $1$ is implemented by the network $Q$ with the $%
\{1\} $-fault for which the constant $1$ is assigned to each edge of $Q$. One
can show that $1\neq x_{1}^{b_{1}}\vee \cdots \vee x_{n}^{b_{n}}$ for any
tuple $(b_{1},\ldots ,b_{n})\in \{0,1\}^{n}$ with exactly $k$ zeros and $\
\binom{n}{\left\lceil \frac{n}{2}\right\rceil }$ is the minimum cardinality
of a set of tuples from $\{0,1\}^{n}$ on which the function $1$ is different
from the functions $x_{1}^{b_{1}}\vee \cdots \vee x_{n}^{b_{n}}$ such that
the tuple $(b_{1},\ldots ,b_{n})\ $belongs to $\{0,1\}^{n}$ and has exactly $%
k$ zeros. It is clear that any $\{1\}$-fault of $Q$ is a $\{0,1\}$-fault of $%
Q$. By Lemma \ref{L8.1.1}, $h_{C}(Q)\geq \binom{n}{\left\lceil \frac{n}{2}%
\right\rceil }$.

(b) Let $C=\{0\}$. We denote by $P$ the switching network with $n$ input
variables $x_{1},\ldots ,x_{n}$, which implements the symmetric function $%
\psi =\psi _{n}^{=\left\lceil \frac{n}{2}\right\rceil }$ and for which $%
L(P)=n^{2}+n$  (see Sect. \ref{S8.1.2}).

Let $Q$ be a switching network with $n$ input variables $x_{1},\ldots ,x_{n}$%
, which implements the function $\psi \,$. Let $\tilde{\delta}=(\delta
_{1},\ldots ,\delta _{n})$ be a tuple from the set $\{0,1\}^{n}$ containing
exactly $k=\left\lceil \frac{n}{2}\right\rceil $ ones. Then $\psi (\tilde{\delta})=1$.  For definiteness, we
assume that $\tilde{\delta}=(1,\ldots ,1,0,\ldots ,0)$. Since $\psi (\tilde{\delta})=1$, in $Q$ there
exists a simple path $\xi $ between poles of $Q$ for which the elementary
conjunction $c_{\lambda }(\xi )$ corresponding to this path is equal to $1$
for the tuple $\tilde{\delta}$, where $\lambda $ is the empty fault of $Q$.
Let us show that $c_{\lambda }(\xi )=x_{1}\cdots x_{k}\bar{x}_{k+1}\cdots
\bar{x}_{n}$.

It is clear, that all variables without negation from $%
c_{\lambda }(\xi )$ belong to the set $A=\{x_{1},\ldots ,x_{k}\}$ and all
variables with negation from $c_{\lambda }(\xi )$ belong to the set $B=\{%
\bar{x}_{k+1},\ldots ,\bar{x}_{n}\}$. Let at least one literal from $A$ do
not belong to $c_{\lambda }(\xi )$. Then $c_{\lambda }(\xi )$ is equal to $1$
on a tuple with less than $k$ ones.
Let at least one literal from $B$ do
not belong to $c_{\lambda }(\xi )$. Then $c_{\lambda }(\xi )$ is equal to $1$
on a tuple with greater than $k$ ones. Thus, $c_{\lambda }(\xi )=x_{1}\cdots
x_{k}\bar{x}_{k+1}\cdots \bar{x}_{n}$.

Let us consider a $\{0\}$-fault of $Q$ that
consists in assigning the constant $0$ to each edge that does not belong to
the path $\xi $. Then the switching network $Q$ with this fault implements
the function $c_{\lambda }(\xi )=x_{1}\cdots x_{k}\bar{x}_{k+1}\cdots \bar{x}%
_{n}$.

Thus, for any tuple $(b_{1},\ldots ,b_{n})\in \{0,1\}^{n}$ with exactly $k$
ones, $x_{1}^{b_{1}}\cdots x_{n}^{b_{n}}$ is the function implemented by the
switching network $Q$ with some $\{0\}$-fault. It is clear that the function
$0$ is implemented by $Q$ with $\{0\}$-fault that consists in assigning the
constant $0$ to each edge of $Q$. One can show that $0\neq
x_{1}^{b_{1}}\cdots x_{n}^{b_{n}}$ for any tuple $(b_{1},\ldots ,b_{n})\in
\{0,1\}^{n}$ with exactly $k$ ones and $\ \binom{n}{\left\lceil \frac{n}{2}%
\right\rceil }$ is the minimum cardinality of a set of tuples from $%
\{0,1\}^{n}$ on which the function $0$ is different from the functions $%
x_{1}^{b_{1}}\cdots x_{n}^{b_{n}}$ such that the tuple $(b_{1},\ldots
,b_{n})\ $belongs to $\{0,1\}^{n}$ and has exactly $k$ ones. By Lemma \ref%
{L8.1.1}, $h_{C}(Q)\geq \binom{n}{\left\lceil \frac{n}{2}\right\rceil }$.
\end{proof}

\begin{remark} \label{R8.1.1}
It is well known that $ \binom{n}{\left\lceil \frac{n}{2}\right\rceil}=\max\left\{ \binom{n}{k}: k=0,\ldots , n\right\}$. Therefore $$\binom{n}{\left\lceil \frac{n}{2}\right\rceil} \ge \frac{2^n}{n+1}\;.$$
\end{remark}

\section{Complexity of Diagnostic Decision Tree Construction \label%
{S8.1.4}}

In this section, we consider one more negative result: we prove that the
problem of construction of decision trees for diagnosis of constant faults
in switching networks is NP-hard.

Let $C\in \{\{0,1\},\{0\},\{1\}\}$. We now define an algorithmic problem $%
Con(C)$: for a given switching network $P$ with $L(P)\geq 1$ and a set $R$
of $C$-faults of $P$, it is required to construct a decision tree $\Gamma $,
which solves the problem of diagnosis of $C$-faults for $P$ and $R$. One can
show that there exists a decision tree, which solves this problem and has at
most $2\left\vert R\right\vert -1$ nodes.

\begin{theorem}
\label{T8.1.3} Let $C\in \{\{0,1\},\{0\},\{1\}\}$. Then the problem $Con(C)$
is NP-hard.
\end{theorem}

\begin{proof}
Let us assume that there exists an algorithm, which solves the problem $%
Con(C)$ and has polynomial time complexity. We now show that in this case
there exists an algorithm, which has polynomial time complexity and solves
the vertex cover problem that is NP-hard. In this problem, for a given
simple undirected graph $G=(V,E)$ with the set of vertices $V$ and the set
of edges $E$, and a natural number $m$, $0<m<\left\vert V\right\vert $, it is
required to recognize if there exists a subset $W$ of the set $V$ such that
each edge from $E$ is incident to at least one node from $W$ and $\left\vert
W\right\vert \leq m$. Let $V=\{v_{1},\ldots ,v_{n}\}$ and $%
E=\{\{v_{i_{1}},v_{j_{1}}\},\ldots ,\{v_{i_{t}},v_{j_{t}}\}\}$. Let us
consider the Boolean function $\psi _{G}(x_{1},\ldots ,x_{n})=(x_{i_{1}}\vee
x_{j_{1}})\cdots (x_{i_{t}}\vee x_{j_{t}})$. It is clear that the vertex cover problem defined above has a solution if and only if there exists a tuple $(\delta
_{1},\ldots ,\delta _{n})\in \{0,1\}^{n}$ with at most $m$ ones such that $%
\psi _{G}(\delta _{1},\ldots ,\delta _{n})=1$. It is also clear that the
considered problem has a solution if and only if $\psi _{n}^{\geq m+1}\neq
\psi _{n}^{\geq m+1}\vee \psi _{G}$ (see definition of the function $\psi _{n}^{\geq m+1}$ in Sect. \ref{S8.1.2}).

\begin{figure}[h]
	\begin{center}
		\includegraphics[width=40mm]{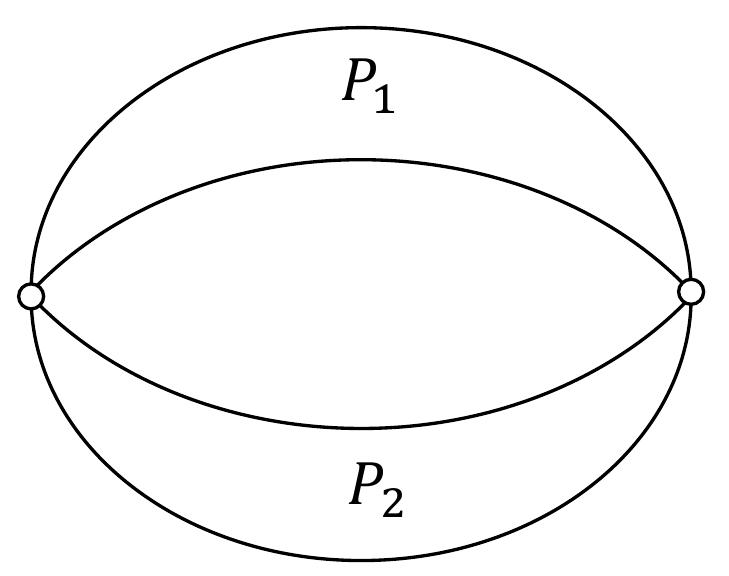}
		\caption{Switching network $Q_1$}
		\label{fig8.4}
	\end{center}
\end{figure}

\begin{figure}[h]
	\begin{center}
		\includegraphics[width=55mm]{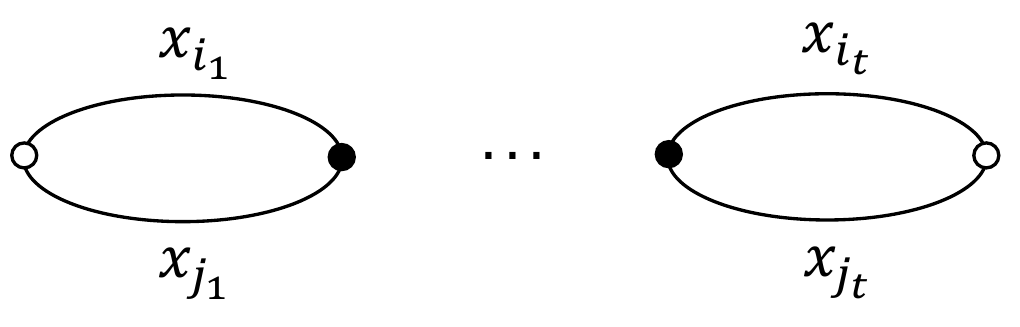}
		\caption{Subnetwork $P_2$}
		\label{fig8.6}
	\end{center}
\end{figure}

(a) Let $C\in \{\{0,1\},\{0\}\}$. It is clear that in a polynomial time
depending on the length of description of the graph $G$ we can construct a
switching network $Q_1$ depicted in Fig. \ref{fig8.4} in which the subnetwork $P_{1}$
implements the function $\psi _{n}^{\geq m+1}$ and has $n^{2}+n$ edges (see
Sect. \ref{S8.1.2}) and the subnetwork $P_{2}$ implements the function $\psi
_{G}$ and has $2t$ edges (see Fig. \ref{fig8.6}). Let $R=\{\lambda ,\rho \}$, where $%
\lambda $ is the empty fault of $Q_1$ and $\rho $ is a fault of $Q_1$ that
consists in assigning the constant $0$ to each edge of the subnetwork $P_{2}$%
. Both faults from $R$ are $\{0\}$-faults and therefore $\{0,1\}$-faults. It
is clear that the network $Q_1$ with the fault $\lambda $ implements the
function $\psi _{n}^{\geq m+1}\vee \psi _{G}$ and the network $Q_1$ with the
fault $\rho $ implements the function $\psi _{n}^{\geq m+1}$.

Apply the
algorithm, which solves the problem $Con(C)$ and has polynomial time
complexity to the network $Q_1$ and the set of faults $R$. As a result, we
obtain a decision tree $\Gamma $, which solves the problem of diagnosis of $%
C $-faults for $Q_1$ and $R$. Apply $\Gamma $ to the network $Q_1$ with the
fault $\lambda $ and apply $\Gamma $ to the network $Q_1$ with the fault $\rho
$. The considered vertex cover problem has a solution if and only if the
results of work of $\Gamma $ for these faults are different. Note that
the processes of the construction of $Q_1$, $\lambda $, $\rho $ and $\Gamma $
and the application of $\Gamma $ to $Q_1$ and faults $\lambda $ and $\rho $
can be done in a polynomial time. Thus, if there is a polynomial time
algorithm for the problem $Con(C)$, then there is a polynomial time
algorithm for the vertex cover problem. Hence the problem $Con(C)$ is NP-hard.

\begin{figure}[h]
	\begin{center}
		\includegraphics[width=45mm]{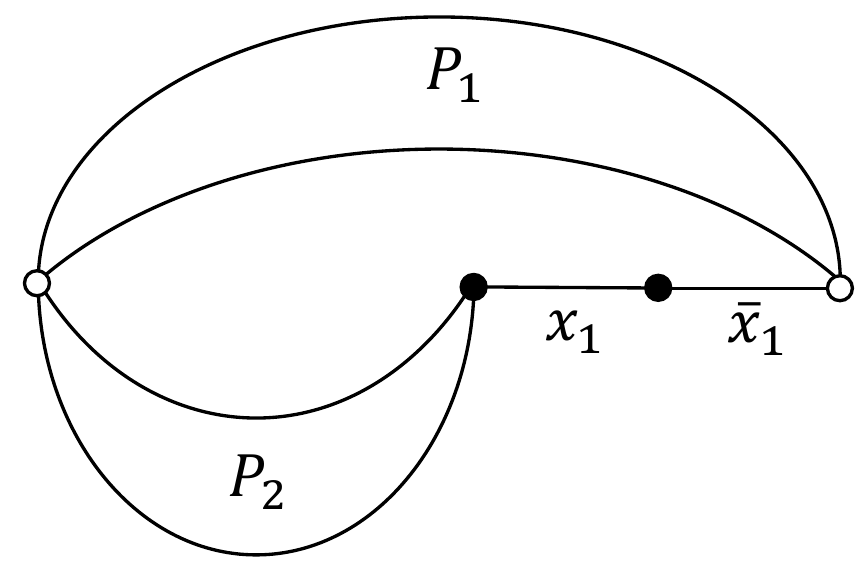}
		\caption{Switching network $Q_2$}
		\label{fig8.5}
	\end{center}
\end{figure}

(b) Let $C=\{1\}$. It is clear that in a polynomial time depending on the length
of description of the graph $G$ we can construct a switching network $Q_2$
depicted in Fig. \ref{fig8.5} in which the subnetwork $P_{1}$ implements the function $%
\psi _{n}^{\geq m+1}$ and has $n^{2}+n$ edges (see Sect.  \ref{S8.1.2}) and the
subnetwork $P_{2}$ implements the function $\psi _{G}$ and has $2t$ edges
(see Fig. \ref{fig8.6}). Let $R=\{\lambda ,\rho \}$, where $\lambda $ is the empty fault
of $Q_2$ and $\rho $ is a $\{1\}$-fault of $Q_2$ that consists in assigning the
constant $1$ to two edges in the right-hand side of the network $Q_2$ that are
labeled with the literals $x_{1}$ and $\bar{x}_{1}$, respectively. It is
clear that the network $Q_2$ with the fault $\lambda $ implements the function
$\psi _{n}^{\geq m+1}$ and the network $Q_2$ with the fault $\rho $ implements
the function $\psi _{n}^{\geq m+1}\vee \psi _{G}$.

Apply the algorithm,
which solves the problem $Con(C)$ and has polynomial time complexity to the
network $Q_2$ and the set of faults $R$. As a result, we obtain a decision
tree $\Gamma $, which solves the problem of diagnosis of $C $-faults for $Q_2$
and $R$.
Apply $\Gamma $ to the network $Q_2$ with the fault $\lambda $ and
apply $\Gamma $ to the network $Q_2$ with the fault $\rho $. The considered
vertex cover problem has a solution if and only if the results of work of $%
\Gamma $ for these faults are different. Note that the processes of the
construction of the switching network $Q_2$, faults $\lambda $, $\rho $ and
decision tree $\Gamma $ and the application of $\Gamma $ to $Q_2$ and
faults $\lambda $ and $\rho $ can be done in a polynomial time. Thus, if
there is a polynomial time algorithm for the problem $Con(C)$, then there is
a polynomial time algorithm for the vertex cover problem. Hence the problem $%
Con(C)$ is NP-hard.
\end{proof}

\section{Conclusions} \label{S8.1.5}

In this paper, we studied decision trees for diagnosis of three types of constant faults in arbitrary switching networks.
We shown that there are simple switching networks for which the depth of decision trees for diagnosis of constant faults is high. We also shown that there are simple switching networks such that, for all switching networks implementing the same functions, the depth of decision
trees for diagnosis of constant faults is high. We proved that the problem of the construction of diagnostic decision trees is NP-hard.

\subsection*{Acknowledgements}
Research reported in this publication was supported by King Abdullah University of Science and Technology (KAUST).
\bibliographystyle{spmpsci}

\bibliography{papers}

\end{document}